# Systematic Procedural and Sensitivity Analysis of the Pattern Informatics Method for Forecasting Large (M > 5) Earthquake Events in Southern California

J. R. Holliday[1,2], J. B. Rundle[1,2], K. F. Tiampo[3], W. Klein[4], and A. Donnellan[5]

[1]Center for Computational Science and Engineering, University of California, One Shields Avenue, Davis, CA 95616-8677, USA.

[2]Department of Physics, University of California, One Shields Avenue, Davis, CA 95616-8677, USA.

[3]Department of Earth Sciences, University of Western Ontario, Biology and Geological Sciences Bldg., London, Ontario, CANADA N6A 5B7.

[4]Department of Physics, Boston University, 590 Commonwealth Avenue, Boston, MA 02215, USA.

[5]Earth and Space Sciences Division, Jet Propulsion Laboratory, Mail Stop 183-335, 4800 Oak Grove Drive, Pasadena, CA 91109-8099, USA.

E-mail: holliday@cse.ucdavis.edu, jbrundle@ucdavis.edu, ktiampo@uwo.ca, klein@buphyc.bu.edu, donnellan@jpl.nasa.gov

Corresponding author:

J.R. Holliday

Center for Computational Science and Engineering

University of California

One Shields Avenue

Davis, CA 95616-8677

USA

E-mail: holliday@cse.ucdavis.edu

Tel: +1-530-752-6419

Fax: +1-530-754-4885

## Abstract

Recent studies in the literature have introduced a new approach to earthquake forecasting based on representing the space-time patterns of localized seismicity by a time-dependent system state vector in a real-valued Hilbert space and deducing information about future space-time fluctuations from the phase angle of the state vector. While the success rate of this Pattern Informatics (PI) method has been encouraging, the method is still in its infancy. Procedural analysis, statistical testing, parameter sensitivity investigation and optimization all still need to be performed. In this paper, we attempt to optimize the PI approach by developing quantitative values for "predictive goodness" and analyzing possible variations in the proposed procedure. In addition, we attempt to quantify the systematic dependence on the quality of the input catalog of historic data and develop methods for combining catalogs from regions of different seismic rates.

## 1. Introduction

Large magnitude earthquakes are devastating events which can have great social, scientific, and economic impact. The 26 December 2003 magnitude 6.7 Iran earthquake killed nearly 30,000 persons. The 16 January 1995 Japan magnitude 6.9 earthquake produced an estimated $200 billion loss. Similar scenarios are possible at any time in San Francisco, Seattle, and other U.S. urban centers along the Pacific plate boundary, especially in Southern California. The gravity of potential loss of life and property is so great that reliable earthquake forecasting should be at the forefront of research goals.

While millions of dollars and thousands of work years have been spent on observational programs searching for reliable precursory phenomena, to date few successes have been reported and no precursors to large earthquake events have been detected that provide reliable forecasts. Indeed, many wonder if earthquake forecasting is even possible (see, for example, the online debate hosted at http://www.nature.com/nature/debates/earthquake).

A new approach to earthquake forecasting, the pattern informatics (PI) approach, has been proposed by *Rundle et al.* (2000a, 2000b, 2002, 2003) and *Tiampo et al.* (2002a, 2002b, 2002c). This approach is based on the strong space-time correlations that are responsible for the cooperative behavior of driven threshold systems and arise both from threshold dynamics as well as from the mean field (long range) nature of the interactions.

Using both simulations and observed earthquake data, they have shown that the space-time patterns of threshold events (earthquakes) can be represented by a time-dependent system state vector in a Hilbert space. The length of the state vector represents the average temporal frequency of events throughout the region and is closely related to the rate at which stress is dissipated. It can be deduced that the information about space-time fluctuations in the system state is represented solely by the phase angle of the state vector. Changes in the norm of the state vector represent only random fluctuations and can for the most part be removed by requiring the system state vector to have a constant norm. A more detailed summary of the method is given in section 4.

## 2. Background

Earthquake fault systems are now believed to be a complex example of a highly nonlinear system (*Bak and Tang*, 1989; *Rundle and Klein*, 1995). Interactions among a spatial network of fault segments are mediated by means of a potential that allows stresses to be redistributed to other segments following slip on any particular segment. For faults embedded in a linear elastic host, this potential is a stress Green's function whose exact form can be calculated from the equations of linear elasticity, once the current geometry of the fault system is specified. A persistent driving force, arising from plate tectonic motions, increases stress on the fault segments. Once the stresses reach a threshold characterizing the limit of stability of the fault, a sudden slip event results. The slipping segment can also trigger slip at other locations on the fault surface whose stress levels are near the failure threshold as the event begins. In this manner, earthquakes occur that result from the interactions and nonlinear nature of the stress thresholds.

The Karhunen-Loeve method (*Fukunaga*, 1970; *Holmes et al.*, 1996), a linear decomposition technique in which a dynamical system is decomposed into a complete set of orthonormal subspaces, has been applied to a number of other complex nonlinear systems over the last fifty years, including the ocean-atmosphere interface, turbulence, meteorology, biometrics, statistics, and even solid earth geophysics (*Hotelling*, 1993; *Fukunaga*, 1970; *Aubrey and Emery*, 1983; *Preisendorfer*, 1988; *Savage*, 1988; *Penland*, 1989; *Vautard and Ghil*, 1989; *Garcia and Penland*, 1991; *Penland and Magorian*, 1993; *Penland and Sardeshmukh*, 1995; *Holmes et al.*, 1996; *Moghaddam et al.*, 1998). The notable success of this method in analyzing the ocean-atmosphere interface and such features as the El Niño Southern Oscillation (ENSO), a nonlinear system whose underlying physics is governed by the Navier-Stokes equation, suggested its application to the analysis of the earthquake fault system (*North*, 1984; *Preisendorfer*, 1988; *Penland and Magorian*, 1993; *Penland and Sardeshmukh*, 1995). Building on these methods for analyzing nonlinear threshold systems, space-time seismicity patterns can be identified in both observed phenomena and numerical simulations using realistic earthquake models for southern California (*Bufe and Varnes*, 1993; *Bowman et al.*, 1998; *Gross and Rundle*, 1998; *Brehm and Braile*, 1999; *Jaume and Sykes*, 1999; *Tiampo et al.*, 1999, 2000; *Rundle et al.*, 2000b.

The PI method is an adaptation of the Karhunen-Loeve expansion technique to the analysis of observed seismicity data from southern California in order to identify basis patterns for all possible space-time seismicity configurations. These basis states represent a complete, orthonormal set of eigenvectors and associated eigenvalues, obtained from the diagonalization of the correlation operators computed for the regional historic seismicity data, and, as such, can be used to reconstitute the data for various subset time periods of the entire data set.

## 3. Data

The primary data set employed in this analysis is the entire historic seismic catalog from 1 January 1932 through 31 December 1999, obtained from the Southern California Earthquake Data Center (SCEDC) online searchable database[1], with all non-local and blast events specifically removed. The relevant data consists of location, in East longitude and North latitude, and the date the event occurred. Seismic events between $-122^{o}$ and $-115^{o}$ longitude and between $32^{o}$ and $37^{o}$ latitude (any depth and quality) and with magnitude greater than or equal to $M_{min} = 3.0$ were selected.

---

[1] http://www.data.scec.org/catalog_search/index.html

While the SCEDC catalog is among the best available, both in completeness and historic depth, there are a number of known deficiencies[2] that undoubtedly affect the quality of our constructed forecast hot-spot maps. The most notable of these issues is that the four-year span of data from 1977-1980 is currently not available to web searching. Fortunately, data for these missing years is available from the older Southern California Seismic Network (SCSN) archives[3] and was hand inserted for this analysis. Unless otherwise indicated, all analysis was performed using SCEDC data with the additional SCSN data.

A second source of data employed in this analysis was acquired from the Northern California Earthquake Data Center (NCEDC) online searchable database[4], with all non-local and blast events again specifically removed. When incorporating this catalog, seismic events between -122$^o$ and -115$^o$ longitude and between 35$^o$ and 37$^o$ latitude (any depth and quality) and with magnitude greater than or equal to $M_{min}$ = 3.0 were selected. The necessity for utilizing an additional catalog in some of our analysis arises from various earthquake events in the vicinity of 35$^o$ North latitude missing from the SCEDC catalog but present in the NCEDC collection.

## 4. Basic Method

Here we summarize the current PI method as described by *Rundle et al.* (2003) and *Tiampo et al.* (2002c). The PI approach is a six step process that creates a time-dependent system state vector in a real valued Hilbert space and uses the phase angle to predict future states (*Rundle et al.*, 2003). The method is based on the idea that the future time evolution of seismicity can be described by pure phase dynamics (*Mori and Kuramoto*, 1998; *Rundle et al.*, 2000a, 2000b). Hence, a real-valued seismic phase function $\hat{S}(x_i, t_b, t)$ is constructed and allowed to rotate in its Hilbert space. Since seismicity in active regions is a noisy function (*Kanamori*, 1981), only temporal averages of seismic activity are utilized in the method. The geographic area of interest is partitioned into $N$ square bins centered on a point $x_i$ and with an edge length $dx$ determined by the nature of the physical system. For our analysis we chose $dx = 0.1°$ ~ 11km, corresponding to the linear size of a magnitude M ~ 6 earthquake. Within each box, a time series $\psi_{obs}(x_i, t)$ is defined by counting how many earthquakes with magnitude greater than $M_{min}$ occurred during the time period $t$ to $t + dt$. Next, the activity rate function $S(x_i, t_b, T)$ is defined as the average rate of occurrence of earthquakes in box $i$ over the period $t_b$ to $T$:

$$S(x_i, t_b, T) = \frac{\sum \psi(x_i, t)}{T - t_b}. \tag{1}$$

---

[2]http://www.data.scec.org/catalog_search/known_issues.html
[3]http://www.data.scec.org/ftp/catalogs/SCSN/
[4]http://quake.geo.berkeley.edu/ncedc/catalog-search.html

If $t_b$ is held to be a fixed time, $S(x_i, t_b, T)$ can be interpreted as the $i$th component of a general, time-dependent vector evolving in an $N$-dimensional space (*Tiampo et al.*, 2002c). Furthermore, it can be shown that this $N$-dimensional correlation space is defined by the eigenvectors of an $N$x$N$ correlation matrix (*Rundle et al.*, 2000a, 2000b). The activity rate function is then normalized by subtracting the spatial mean over all boxes and scaling to give a unit-norm:

$$\hat{S}(x_i, t_b, T) = \frac{S(x_i, t_b, T) - \frac{1}{N} \sum S(x_j, t_b, T)}{\sqrt{\sum \left[ S(x_j, t_b, T) - \frac{1}{N} \sum S(x_k, t_b, T) \right]^2}}. \tag{2}$$

The requirement that the rate functions have a constant norm helps remove random fluctuations from the system. Following the assumption of pure phase dynamics (*Rundle et al.*, 2000a, 2000b), the important changes in seismicity will be given by the change in the normalized activity rate function for the time period $t_1$ to $t_2$:

$$\Delta \hat{S}(x_i, t_b, t_1, t_2) = \hat{S}(x_i, t_b, t_2) - \hat{S}(x_i, t_b, t_1) . \tag{3}$$

This is simply a pure rotation of the *N*-dimensional unit vector $\hat{S}(x_i, t_b, T)$ through time. In order to remove the last free parameter in the system, the choice of base year, and to further reduce random noise components, changes in the normalized activity rate function are averaged over all possible base-time periods:

$$\Delta \underline{\hat{S}}(x_i, t_0, t_1, t_2) = \frac{\sum_{t_b = t_0}^{t_1} \Delta \hat{S}(x_i, t_b, t_1, t_2)}{t_1 - t_0} . \tag{4}$$

Finally, the probability of change of activity in a given box is deduced from the square of its base averaged, mean normalized change in activity rate:

$$P(x_i, t_0, t_1, t_2) = [\Delta \underline{\hat{S}}(x_i, t_b, t_1, t_2)]^2 . \tag{5}$$

In phase dynamical systems, probabilities are related to the square of the associated vector phase function (*Mori and Kuramoto*, 1998; *Rundle et al.*, 2000b). This probability function is often given relative to the background by subtracting off its spatial mean:

$$P'(x_i, t_0, t_1, t_2) = P(x_i, t_0, t_1, t_2) - \frac{1}{N} \sum P(x_j, t_0, t_1, t_2), \tag{6}$$

where $P'$ indicates the probability of change in activity and is measured relative to the background.

Schematically, this whole process can be represented by

$$N \to S \to \hat{S} \to \Delta \hat{S} \to \Delta \underline{\hat{S}} \to P,$$

where the *hat* symbol is understood to mean "calculate normalization in space", the capital Delta means "calculate the change in rate", and the underscore symbol means "average over base times". Note that this method implicitly assumes earthquake fault systems are in an unstable equilibrium state and can be treated linearly about their equilibrium points.

## 4.1. Variations in Order

To determine the optimal application of the PI method, we identified and analyzed all physically meaningful variations of the described procedure. While we have outlined above a six step process, there are considerably fewer than 6! = 720 variations that need to be investigated. A forecast analysis must always begin with binning the available data and end with a calculation of probability change. Also, base-time averaging and calculation of changes in the activity rate functions can only be performed after creating the activity rate vectors. With these constraints imposed, there are only eight possible variations in the order to which each step is performed. Table 1 lists these eight variations with the original method denoted Method **I**.

On the basis of theoretical arguments and assumptions of linearity within the system, we expect that Methods **I** through **VI** should perform qualitatively similar to each other. This is due largely to the fact that the operations being permuted are all linear and commute with each other. Qualitatively it is unclear which variation should yield the best correlation with actual future events other than to expect Methods **II** and **III** might perform better than Method **I** due to the movement of when the change in activity rate is calculated to after the normalization and base-time averaging steps. This essentially places all of the activity rate vectors on equal footing and legitimizes the vector rotation. We also expect that Methods **VII** and **VIII** will yield both qualitatively and quantitatively inferior forecast hot-spot maps. This is due to the direct normalization of the binned data. Such a step destroys correlations between different spatial locations by independently scaling the relative historic intensity rates. Each of these expectations are verified in the results section below.

*4.2. Variations in Binning*

In addition to the original binning method, we also analyzed time-centered, cumulative, and detrended binning. For time-centered binning, we took each time series and removed the temporal mean:

$$\psi_{obs}(x_i,t) \rightarrow \psi_{obs}(x_i,t) - \frac{\sum_{t=t_0}^{t_2} \psi_{obs}(x_i,t)}{t_2-t_0} . \tag{7}$$

For cumulative binning we allowed each time series to build on its past events:

$$\psi_{obs}(x_i,t) \rightarrow \sum_{T=t_0}^{t} \psi_{obs}(x_i,T) . \tag{8}$$

For detrended binning, we took each cumulative time series, fit it to a first order polynomial, and subtracted the fitted line:

$$\psi_{obs}(x_i,t) \rightarrow \sum_{T=t_0}^{t} \psi_{obs}(x_i,T) - [A+Bt], \tag{9}$$

where *A* and *B* are the parameters of the regression fit. Figure 1 shows the effect of each binning procedure on a synthetic data sample. We will denote the four different binning methods with the labels **A**, **B**, **C**, and **D**, respectively, with **A** denoting the unmodified method. Methods **B** and **D** are significant in that they remove the mean for each time series from the data. Thus, anomalous activity away from background seismicity is expected to be emphasized. Method **C** is reminiscent of an unbiased estimator in the cumulative distribution Kolmogorov-Smirnov Test (*Press et al.*, 2002) and could in theory allow more accurate comparisons among the different time series.

We also investigated magnitude- and energy-weighted binning where the value at each time step is proportional to either the total magnitude $M_{tot}$ of all the events in the time period or to the total energy ($\sim 10^{Mtot}$) of all the events. These weighting factors, however, had the effect of selecting out time periods surrounding only the largest events and were thus unsuitable for the analysis. We did not investigate *Boolean* binning where each time step is given an initial value of either *1* if one or more events occur in that time period or *0* otherwise due to the realization that this effect can be achieved by sufficiently reducing the time step *dt*. Also, we desired the method to scale appropriately as *dt* is increased.

## *4.3. Variations in Projection*

In addition to calculating the change in the activity rate function through the vector rotation during the time period $t_1$ to $t_2$, we also investigated the effect of linear projection of change into future times:

$$\Delta S(x_i, t_b, t_1, t_2) \rightarrow S(x_i, t_b, t_2) + \Delta S(x_i, t_b, t_1, t_2) \,. \tag{10}$$

The motivation behind this investigation was that for regions with a near constant rate of seismicity (or with frequencies higher than an inverse time step), $\Delta \hat{S}(x_i, t_b, t_1, t_2) \approx 0$. By linear projection, we mean that the future seismic activity for this type of situation would be approximately equal to the present seismic activity with a small correction added. For notational purposes, we will denote the unmodified approach of calculating the change in the activity rate function with the label **1** after the method specification. We will denote the linear projection approach with the label **2**.

*4.4. Variations in dt*

While the spatial width of the boxes, *dx,* is determined by the nature of the physical system, the temporal binning width *dt* is arbitrary. Larger values of *dt* result in greater bin statistics and faster execution time of the algorithm while lower values may potentially yield greater sensitivity to high frequency periodicity.

To investigate the effect, we performed the analysis with representative values for *dt* ranging from one day to one year. If the catalog is uniform in its completeness and not missing bands of data at quasi- periodic intervals, we would expect to find a smooth transition through the varying choices of *dt* with perhaps some optimal selection. On the other hand, large fluctuations in the forecast as *dt* is slowly modified may indicate underlying chaotic phenomena and would bring into question the assumptions and treatment of linearity within the system.

## 5. Statistical Tests

To test the hypothesis that the probability measure $P_i$ can forecast future $(t > t_2)$ large $(M > 5)$ events, we performed a set of maximum likelihood tests [*Bevington and Robinson*, 1992; *Gross and Rundle*, 1998; *Kagan and Jackson*, 2000; *Tiampo et al.*, 2002b; *Schorlemmer et al.*, 2003]. The likelihood **L** is a probability measure that can be used to assess the quality of one forecast measure over another. Typically, one computes $L = \log(\mathbf{L})$ for the proposed forecast measure **L** and compares that to the likelihood measure $L^0 = \log(\mathbf{L^0})$ for a representative null hypothesis. The ratio of these two values then yields information about which measure is more accurate in forecasting future events. In the likelihood ratio test, a probability density function (PDF) is required. Two different PDFs were used in this analysis: a global, Gaussian model and a local, Poissonian model. These distributions differ significantly in that the Gaussian model assumes purely random, normal statistics while the Poissonian model assumes independent statistics over small time intervals with no temporal clustering [*Walpole and Myers*, 1993].

## 5.1. Global Gaussian Model

In their original analysis, *Tiampo et al.* (2002b) calculated likelihood values by defining $P_i = P[x_i]$ to be the union of a set of N Gaussian density functions $p_G(|x-x_i|)$ *(Bevington and Robinson*, 1992*)* centered at each location $x_i$. Each individual Gaussian density has a standard deviation equal to the box width $dx$ and a peak value equal to the calculated probability of change in activity $P_i$ divided by the standard deviation squared. $P[x(e_j)]$ is therefore a probability measure that a future large event $e_j$ occurs at location $x(e_j)$:

$$P[x(e_j)] = \sum_i \frac{P_i}{\sigma^2} e^{\frac{[x(e_j)-x_i]^2}{\sigma^2}} . \tag{11}$$

If there are *J* future events, the normalized likelihood **L** that all *J* events are forecast is:

$$L = \prod_j \frac{P[e(x_j)]}{\sum_i P[x_i]} . \tag{12}$$

Furthermore, the log-likelihood value *L* for a given calculation can be calculated and used in ratio comparison tests:

$$\log(L) = \sum_j \log \frac{P[e(x_j)]}{\sum_i P[x_i]} . \tag{13}$$

Before performing the statistical analysis, the change in activity values $P_i$ were first truncated by scaling all the probabilities equally up-wards and performing a *histogram cut* to enforce the restriction $\Delta P \leq 1$. This was used to eliminate the exponential tail on the high end of the PDF and ensure that events that occurred during the forecasting time period had a probability $\Delta P = 1$ of occurring (which, in fact, they did).

## 5.2. Local Poissonian Model

The second model used is based on work performed by the Regional Earthquake Likelihood Models (RELM) group (*Schorlemmer et al.*, 2003). For each bin $i$ an expectation value $\lambda_i$ is calculated by scaling the local probability $P_i$ by the number of earthquakes that occurred over all space during the forecast time period:

$$\lambda_i = n P_i, \tag{14}$$

where $n$ is the number of post-$t_2$ events. Note that for any future time interval ($t_2$, $t_3$), $n$ could in principle be estimated by using the Gutenberg-Richter relation. For each bin an observation value $w_i$ is also calculated such that $w_i$ contains the number of post-$t_2$ earthquakes that actually occurred in bin $i$. For the RELM model, it is assumed that earthquakes are independent of each other. Thus, the probability of observing $w_i$ events in bin $i$ with expectation $\lambda_i$ is the Poissonian probability

$$p_i(w_i, \lambda_i) = \frac{\lambda_i^{w_i}}{w_i!} e^{-\lambda_i}. \tag{15}$$

The log-likelihood for observing $w$ earthquakes at a given expectation $\lambda$ is defined as the logarithm of the probability $p_i(w_i, \lambda_i)$, thus

$$\log(L(w,\lambda)) = \log(p(w,\lambda)) = -\lambda + w\log(\lambda) - \log(w!). \tag{16}$$

Since the joint probability is the product of the individual bin probabilities, the log-likelihood value for a given calculation is the sum of $\log(L(w,\lambda))$ over all bins $i$.

When using this PDF function, we preprocess the change in activity values $P_i$ by performing the same *histogram cut* as with the Gaussian model.

# 6. Results

Results for the procedural analysis with variations in binning and calculation of activity rate are presented in tables 2 and 3. All values of $L$ are given relative to $L^0$ defined to be the value supplied by our original, unaltered Method **I-A1**. Since these are ratio tests, greater values indicate better predictive ability.

As statistical evaluations of earthquake forecasts are still under development, it is instructive to weigh the quantitative ("predictive goodness" values) against the qualitative (pictorial representation of the forecast hot-spot maps). Thus, representative maps for each procedural variation are given in figures 2 and 3.

Only Methods **II** and **III**, using normal binning and change of activity calculation, performed better than the original method under the two statistical tests. Naively, this result is expected as both methods wait until after normalization and base year averaging to calculate the change in activity rate, thus giving the calculations in each box equal statistical weight. For all other investigated variations, no method performed better on both likelihood tests and qualitative analysis.

While a few of the binning and change of activity variations fared well on one or the other likelihood tests (for example, **III-B1**), most performed poorly qualitatively. Probability calculations gave predictions of activity that spread well into areas with no recorded activity. These results can be understood by considering their mathematical operations. By linearly projecting the change in activity rate, heavy weight is placed on the most recent seismic history. For the procedure to identify anomalous changes in the seismicity, however, the entire history must be considered equally. Also, the cumulative and detrended variations in the binning method create time series that are significantly altered from those apparent in nature.

While only Methods **II-A1** and **III-A1** performed better than the original PI procedure on both statistical tests, it should be stressed that at this time none of the methods can be claimed to be superior. There is still a subjective element over which forecast hot-spot map to prefer. Based on theoretical and mathematical considerations, Method **III-A1** is the authors' preferred choice. This method creates a unique state vector at every time step and allows the purest interpretation of a vector rotation.

Table 4 shows the results of varying the time step in the analysis (note that Method **III-A1** was used). Likelihood values for this investigation were referenced against a choice of $dt$ = 1 day. Note that the accuracy of the calculated forecast decreases with increasing time step, slowly decreasing up to around $dt$ = 1 week and then rapidly decreasing. While larger choices of $dt$ decrease time of computation for the PI algorithm, they do so at the cost of accuracy. Evaluating the data from Table 4, along with the corresponding forecast hot-spot maps, the authors believe $dt$ = 7 days to be a suitable compromise. This choice of time step is low enough to probe the seismic periodicity at all scales with reasonable accuracy while being large enough to significantly speed up the computation.

## 7. Catalog Sensitivity

To gauge the sensitivity of the PI method on the quality of the input catalog, we decimated the available data by systematically increasing both the starting date of catalog information (and thus affecting $t_0$) and the minimum magnitude threshold. Figures 4 and 5 show the effect on the relative likelihood values of varying either parameter individually. Both probability density functions–Poissonian and Gaussian– were used to calculate log likelihood indexes.

In Figure 4 we see the surprising result that the forecast is relatively stable as $t_0$ is increased, up to around 1965. This would indicate that accurate forecast hot-spot maps can be created using only approximately 40 years of historic data. When the normalized activity rate functions are averaged over all possible base-time periods, more recent data gets weighted heavier than more historic data. The threshold for when historic data no longer influences the forecast appears to be approximately 40 years before the onset of the forecast, i.e., $t_2$. With less than 40 years of historic data, however, the likelihood values drop sharply.

The Poissonian analysis in Figure 5 seems to indicate that higher accuracy in the forecast can be obtained by raising the minimum magnitude cut-off threshold of the analysis from $M_{min}$ = 3.0 to ~3.7. This may have the effect of removing low magnitude events that are uncorrelated with future large magnitude events and thereby eliminate background noise from the analysis. Care must be taken, however, as the likelihood values drop quickly as the magnitude threshold is raised too high. It is interesting to note the sudden drop in likelihood values as the magnitude threshold reaches 4.5 (and again near 4.8, 5.5, and 6.3). While statistics may be playing a role in the latter three drops, the discontinuity at $M_{min}$ = 4.5 appears to identify an unknown deficiency in the catalog.

Figures 6 and 7 show the effect on the relative likelihood values of varying both parameters simultaneously. For these two-dimensional plots, warmer colors indicate better correlation between the forecast and actual events. All of the features mentioned above are again evident as well as the surprising observation that increasing $M_{min}$ allows accurate forecasts with less historic data (as indicated by the positive slope of the high-likelihood-edge surrounding $M_{min}$ = 3.6 and $t_0$ = 1967).

## 8. Application Of The Method

To test the our optimization on the PI method, we recreated the forecast seismic hot-spot map originally presented by *Rundle et al.* (2002) for the time period 1 January 2000 to 31 December 2009 using Method **III-A1**. The result is shown in Figure 8. The original forecast was made using only data from the SCEDC catalog, which does not contain earthquakes from the San Simeon region (location of the M=6.5, 2003 event; label #7 in Figure 8). Our revised forecast was made using data from both the NCEDC catalog (for latitude above 35º) and the SCEDC catalog (for latitude below 35º).

Since the cut-off date for the forecast of 31 December 1999, eight large earthquake events with M>5 have occurred in central or southern California. The first seven events all occurred either on areas of forecasted anomalous activity or within the margin of error of +/- 11km. While this hot-spot map was made after each of these events occurred, it was done so using only data prior to 31 December 1999 and could have in principle predicted these events. Scorecards using the original method and the current optimized method can be found at the JPL QuakeSim website[5].

## 9. Combining Catalogs

The issue of how to combine historic catalogs in order to create forecast hot-spot maps for large regions is a difficult one. Problems arise from the fact that different areas will normally have widely different seismic rates, and these differences get smoothed out when we normalize our state vectors.

One way to try and account for these differences is to apply a weighting factor to the different catalogs as they are merged into an aggregate catalog. This method, however, tends to emphasize near threshold-level anomalous activity in the catalog with the highest weighted activity rate. In Figure 9 we created a forecast hot-spot map by combining data from the NCEDC and SCEDC catalogs with two different weighting ratios. With equal weighting between the two catalogs (Figure 9A), event #3 (Anza) occurs near a threshold-level anomalous region. Event #7 (San Simeon), however, is missed completely. As the relative weighting for the northern catalog is increased to account for its lower total seismic rate (Figure 9B), anomalous activity begins to appear under event #7, but disappears from event #3.

Another way to try and account for the differences is to apply a weighting factor to each individual time series based on its own statistics. This method, unfortunately, also has failings. By weighing each time series individually, correlations between local events are destroyed. In practice, this approach has effects similar to the earlier proposed modifications **VII** and **VIII** to the PI procedure and simply results in more apparent noise in the forecast and less correlation with actual future events.

---

[5]http://www-aig.jpl.nasa.gov/public/dus/quakesim/scorecard.html

Currently, the best approach (at least for this time period and these catalogs) appears to be to treat all catalogs and regions separately, combining only at the end of the analysis and normalizing over all spatial bins to allow for correlations across the catalog seams.

## 10. Conclusion

We have analyzed the current PI procedure and developed a more optimized approach for creating accurate forecast hot-spot maps. First, historic seismic data is binned by counting the number of earthquakes per unit time, of any size greater than or equal to $M_{min}$, within a geographic box centered at $x_i$ at some time $t$. The geographic region defined by $dx$ is taken large enough so that seismic activity can be considered an incoherent superposition of phase functions. Second, an activity rate function is defined as the average rate of occurrence of earthquakes in box $i$ over the period $t_b$ to $T$. Third, the activity rate function is averaged over all possible base-time periods. Forth, the base-year averaged activity rate function is normalized by subtracting the spatial mean over all boxes and scaling to give a unit-norm. Fifth, changes in the base-year averaged, mean-normalized activity rate function are calculated by allowing the vector to rotate over time. Finally, the probability of change of activity in a given box–calculated relative to the background–is deduced from the square of its base-year averaged, mean-normalized change in activity rate.

We also showed that the choice of *dt* is relatively unimportant to the calculation if it is taken low enough, that only approximately 40 years of complete historic data is necessary for accurate forecasts, and that the assumptions of linearity and near-equilibrium appear valid for Southern California seismic fault systems. Applying our new procedure, we recalculated and updated the southern California forecast hot-spot map presented by *Rundle et al.* (2002) and showed that the 22 December 2003 San Simeon event could have been foreseen. Finally, we identified pitfalls associated with combining seismic catalogs from different regions in an attempt to create a composite forecast hot-spot map.

There is movement in the forecast verification community to part with likelihood calculations, which lightly reward successes and heavily penalize failures, and embrace ROC verification diagrams (*Joliffee and Stephenson*, 2003). Additional analyses that utilize these verification techniques are currently underway.

## Acknowledgments

The authors are grateful to the anonymous reviewers for their helpful criticisms and suggestions. This work has been supported by a grant from US Department of Energy, Office of Basic Energy Sciences to the University of California, Davis DE-FG03-95ER14499 (JRH and JBR), by a NASA Earth Science Fellowship NN-6046Q98H (JRH), and through additional funding from the National Aeronautics and Space Administration under grants through the Jet Propulsion Laboratory to the University of California, Davis.

Table 1: Possible variations in the procedure ordering. The analysis must always begin with data binning and end with probability calculation. Recall $N$ is binned data, $S$ is the activity rate, $P$ is a probability calculation, the ˆ symbol represents normalization in space, the $\Delta$ symbol represents calculation of change in rate, and the underscore symbol represents averaging over base times.

| Method | Procedure | | | | | | | | | | |
|---|---|---|---|---|---|---|---|---|---|---|---|
| **I**    | $N$ | $\rightarrow$ | $S$ | $\rightarrow$ | $\hat{S}$         | $\rightarrow$ | $\Delta\hat{S}$         | $\rightarrow$ | $\Delta\underline{\hat{S}}$ | $\rightarrow$ | $P$ |
| **II**   | $N$ | $\rightarrow$ | $S$ | $\rightarrow$ | $\hat{S}$         | $\rightarrow$ | $\underline{\hat{S}}$   | $\rightarrow$ | $\Delta\underline{\hat{S}}$ | $\rightarrow$ | $P$ |
| **III**  | $N$ | $\rightarrow$ | $S$ | $\rightarrow$ | $\underline{S}$   | $\rightarrow$ | $\underline{\hat{S}}$   | $\rightarrow$ | $\Delta\underline{\hat{S}}$ | $\rightarrow$ | $P$ |
| **IV**   | $N$ | $\rightarrow$ | $S$ | $\rightarrow$ | $\Delta S$        | $\rightarrow$ | $\Delta\hat{S}$         | $\rightarrow$ | $\Delta\underline{\hat{S}}$ | $\rightarrow$ | $P$ |
| **V**    | $N$ | $\rightarrow$ | $S$ | $\rightarrow$ | $\Delta S$        | $\rightarrow$ | $\Delta\underline{S}$   | $\rightarrow$ | $\Delta\underline{\hat{S}}$ | $\rightarrow$ | $P$ |
| **VI**   | $N$ | $\rightarrow$ | $S$ | $\rightarrow$ | $\underline{S}$   | $\rightarrow$ | $\Delta\underline{S}$   | $\rightarrow$ | $\Delta\underline{\hat{S}}$ | $\rightarrow$ | $P$ |
| **VII**  | $N$ | $\rightarrow$ | $\check{N}$ | $\rightarrow$ | $\hat{S}$   | $\rightarrow$ | $\Delta\hat{S}$         | $\rightarrow$ | $\Delta\underline{\hat{S}}$ | $\rightarrow$ | $P$ |
| **VIII** | $N$ | $\rightarrow$ | $\check{N}$ | $\rightarrow$ | $\hat{S}$   | $\rightarrow$ | $\underline{\hat{S}}$   | $\rightarrow$ | $\Delta\underline{\hat{S}}$ | $\rightarrow$ | $P$ |

Table 2: Relative likelihood values $L_G - L^0$ using a global Gaussian model over the time period t = 1984 → 1994 for the various variations in order, binning, and calculation of change in activity rate. Recall that **A** – **D** denote normal, time-centered, cumulative, and detrended binning, respectively, while **1** and **2** denote normal and projected calculations of change in activity rate. For our null hypothesis, $L^0$, we took the value from Method **I-A1**. Larger (more positive) values are better correlated with actual events.

| Method | A1 | B1 | C1 | D1 | A2 | B2 | C2 | D2 |
|---|---|---|---|---|---|---|---|---|
| I | 0.00 | -13.06 | -11.27 | -18.80 | -36.47 | -32.23 | -19.43 | -24.62 |
| II | 3.33 | -8.65 | -21.91 | -17.96 | -36.14 | -30.92 | -14.17 | -23.27 |
| III | 2.70 | -1.04 | -32.58 | -19.89 | -15.28 | -15.28 | -14.74 | -21.99 |
| IV | -2.89 | -2.08 | -16.10 | -13.87 | -31.20 | -16.43 | -15.94 | -12.57 |
| V | -7.99 | -4.75 | -14.35 | -19.70 | -34.48 | -12.94 | -14.67 | -21.51 |
| VI | -2.76 | -2.92 | -17.63 | -19.92 | -33.23 | -10.88 | -14.54 | -21.05 |
| VII | -20.32 | -17.41 | -14.87 | -32.44 | -48.93 | -10.90 | -16.03 | -33.38 |
| VIII | -16.65 | -21.57 | -37.77 | -32.02 | -47.32 | -10.99 | -15.05 | -33.42 |

Table 3: Relative likelihood values $L_P - L^0$ using a local Poissonian model over the time period t = 1984 → 1994 for the various variations in order, binning, and calculation of change in activity rate. Recall that **A – D** denote normal, time-centered, cumulative, and detrended binning, respectively, while **1** and **2** denote normal and projected calculations of change in activity rate. For our null hypothesis, $L^0$, we took the value from Method **I-A1**. Larger (more positive) values are better correlated with actual events.

| Method | A1 | B1 | C1 | D1 | A2 | B2 | C2 | D2 |
|---|---|---|---|---|---|---|---|---|
| I | -0.00 | 1.29 | -38.14 | -30.87 | -57.74 | -44.65 | -5.77 | -74.67 |
| II | 4.93 | 5.58 | -60.65 | -28.60 | -18.05 | -29.54 | -2.09 | -48.88 |
| III | 2.94 | 14.74 | -59.22 | -26.22 | 5.04 | 5.04 | -2.01 | -35.93 |
| IV | 7.75 | 6.77 | -7.27 | -12.30 | -32.11 | -14.98 | -3.15 | -11.46 |
| V | 0.43 | -0.52 | -7.38 | -43.10 | -45.47 | -5.94 | -2.12 | -45.89 |
| VI | 0.84 | 0.63 | -9.89 | -40.51 | -21.67 | -3.99 | -2.04 | -55.60 |
| VII | -59.34 | -51.33 | -61.89 | -85.76 | -81.90 | -47.86 | -44.11 | -81.12 |
| VIII | -45.73 | -57.16 | -76.22 | -87.33 | -83.09 | -48.66 | -44.12 | -81.55 |

Table 4: Relative likelihood values using Method **III-A1** with varying time steps (in days) over the time period t = 1984 → 1994. For our null hypothesis, we took the value at *dt* = 1 day. Larger (more positive) values are better correlated with actual events.

| $dt$ = | 1 | 3 | 5 | 7 | 15 | 30 | 60 | 90 | 180 | 365 |
|---|---|---|---|---|---|---|---|---|---|---|
| $L_G - L^0$ = | 0.00 | -0.07 | -0.16 | -0.13 | -0.63 | -1.33 | -2.69 | -17.00 | -34.06 | -20.17 |
| $L_P - L^0$ = | 0.00 | -0.55 | -0.70 | -1.43 | -4.52 | -7.31 | -9.66 | -24.43 | -85.22 | -33.66 |

Figure 1: The topmost plot represents random earthquake events over an arbitrary time scale. The four lower plots show the results of the different binning methods: A) normal, B) time-centered, C) cumulative, and D) detrended.

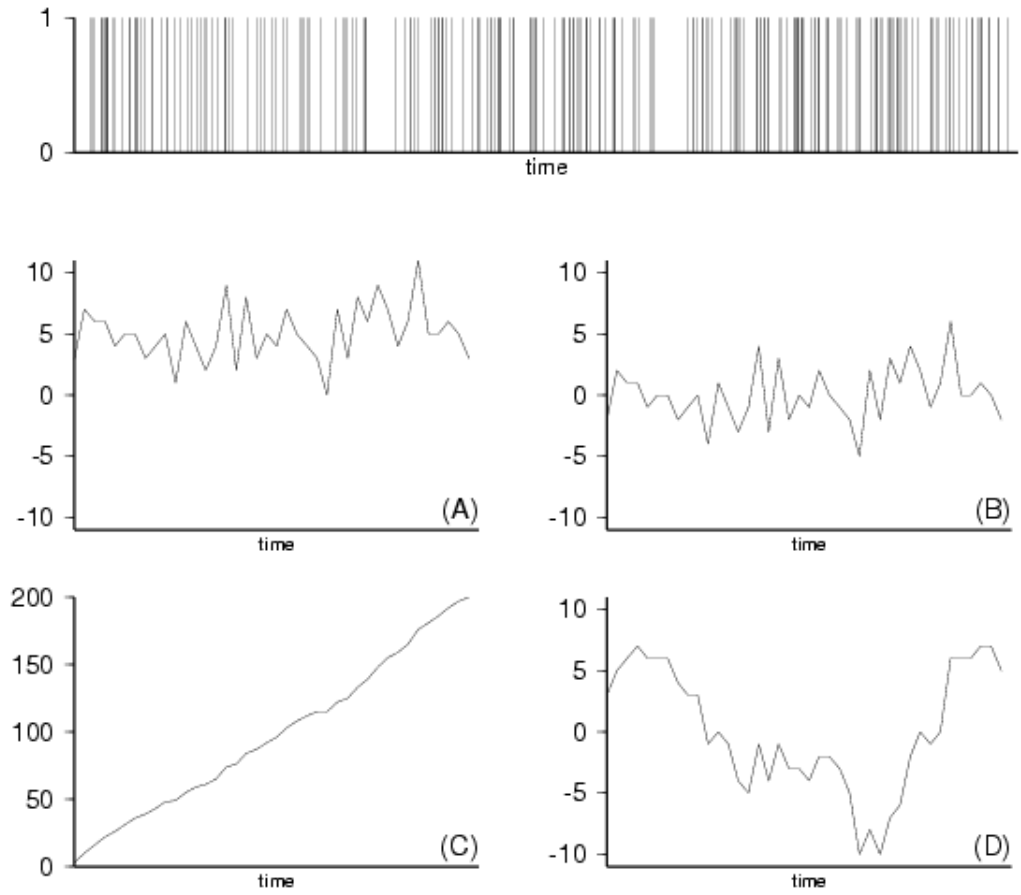

Figure 2: Representative forecast hot-spot maps created using each of the order variations with normal binning and calculation of change in activity rate for the time period t = 1984 to 1994. Note the increase in apparent noise for Methods VII and VIII.

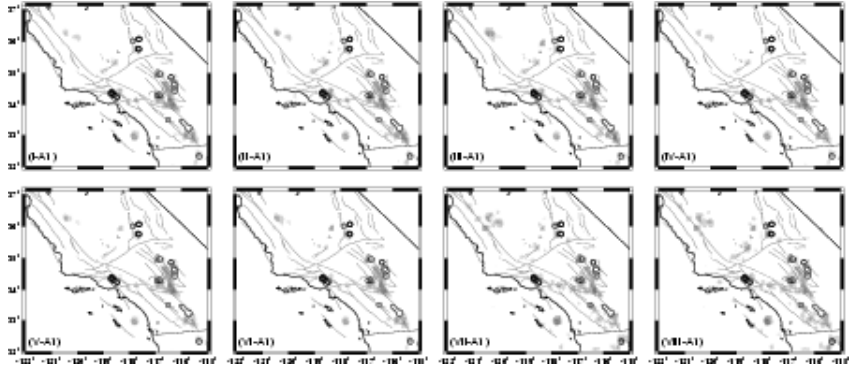

Figure 3: Representative forecast hot-spot maps created using each of the variations in binning and calculation of change in activity rate for Method I over the time period t = 1984 to 1994.

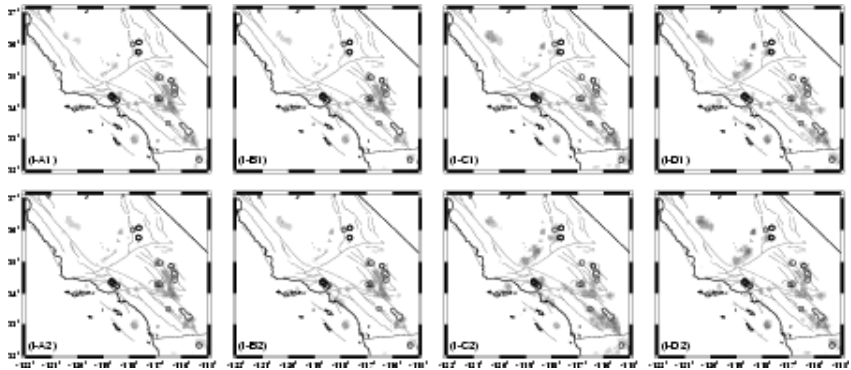

Figure 4: Relative likelihood values for two different probability density functions, Gaussian (solid) and Poissonian (dashed), as a function of $t_0$. Larger (more positive) values are better correlated with actual events. The plateau in the data before $t_0$ = 1965 indicates that only ~40 years of historic data is necessary for the analysis.

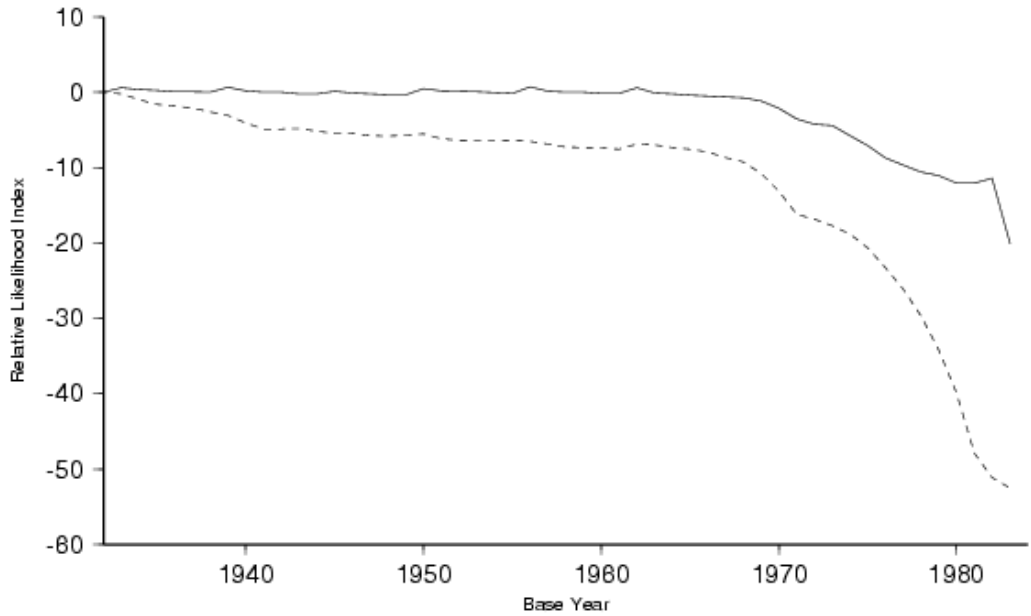

Figure 5: Relative likelihood values for two different probability density functions, Gaussian (solid) and Poissonian (dashed), as a function of the minimum magnitude cut-off threshold. Larger (more positive) values are better correlated with actual events. Using the Poissonian PDF, more probable forecasts appear possible by increasing the magnitude threshold slightly.

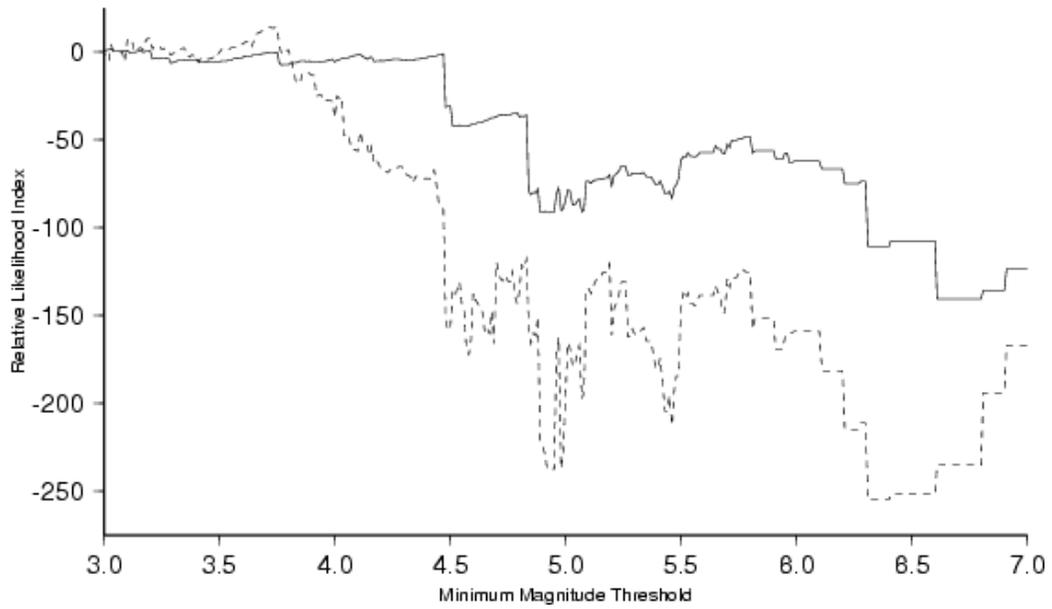

Figure 6: Relative likelihood index calculated using a Gaussian density function as a function of both $t_0$ and minimum magnitude cut-off threshold. Warmer colors are better correlated with actual events.

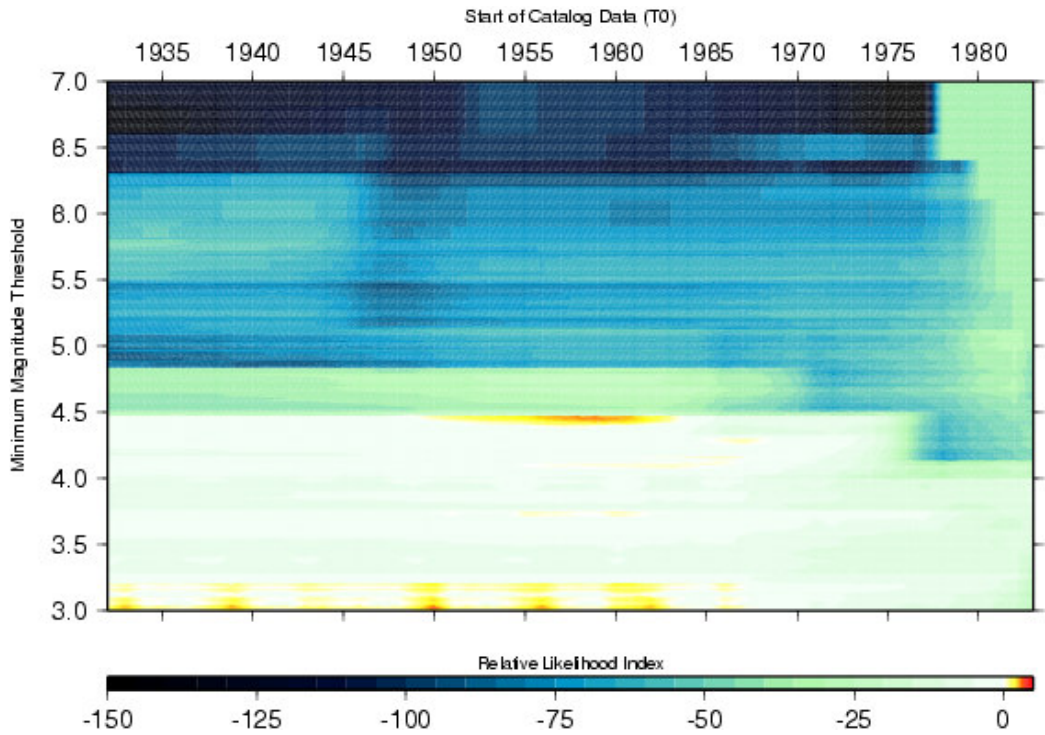

Figure 7: Relative likelihood index calculated using a Poissonian density function as a function of both t₀ and minimum magnitude cut-off threshold. Warmer colors are better correlated with actual events.

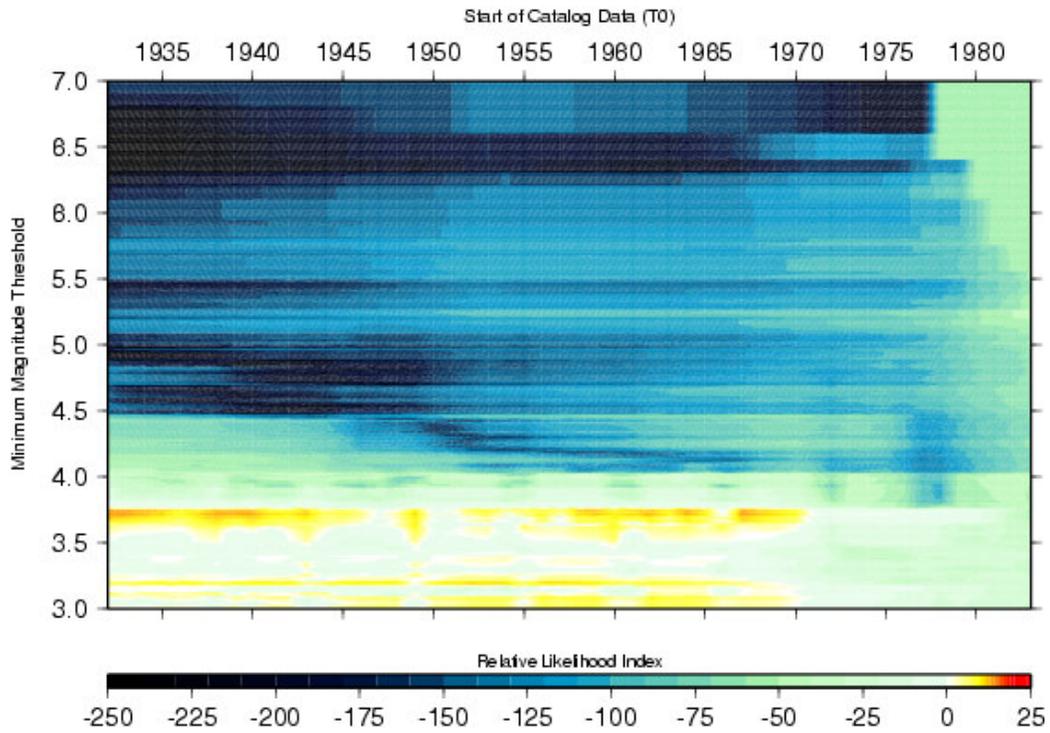

Figure 8: Seismic hot-spot map for large earthquake events with M >5 for the forecast time period 1 January 2000 to 31 December 2009. Since the cut-off date for the forecast, eight large earthquake events with M >5 have occurred in central or southern California. Seven of the eight events occurred either on areas of forecasted anomalous activity or within the margin of error of ±11km. Data from the SCEDC catalog was used below 35 ° North latitude, and from the NCEDC catalog above 35 ° North latitude.

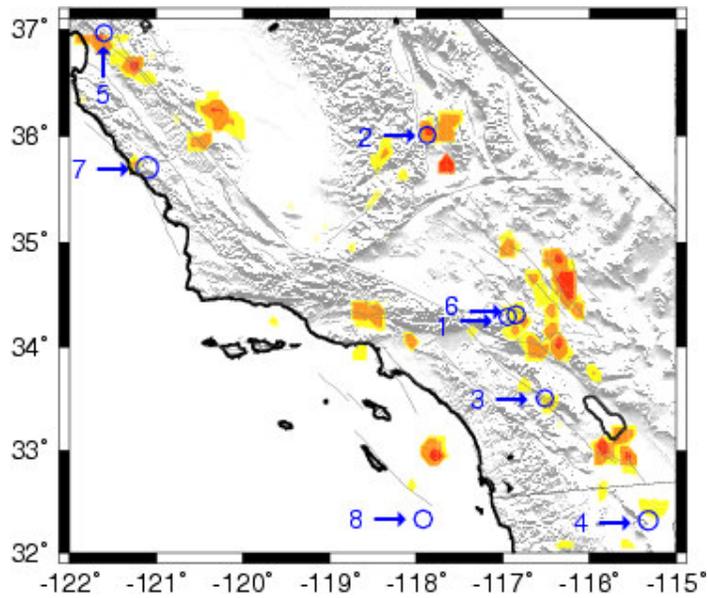

Figure 9: Equal weight for both catalogs (A) vs. higher weighting for northern catalog (B). With the equally weight map, event #3 occurs near a threshold-level anomalous region while event #7 does not. The opposite is true with the unequally weight map.

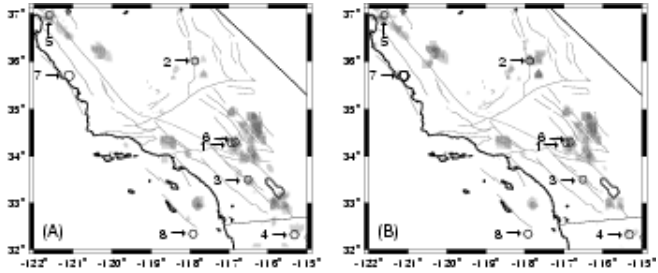